# Sub-cycle optical control of current in a semiconductor: from the multiphoton to the tunneling regime

Tim Paasch-Colberg[1], Stanislav Yu. Kruchinin[1], Özge Sağlam[2], Stefan Kapser[1], Stefano Cabrini[3], Sascha Muehlbrandt[1], Joachim Reichert[2], Johannes V. Barth[2], Ralph Ernstorfer[4], Reinhard Kienberger[2], Vladislav S. Yakovlev[1,5], Nicholas Karpowicz[1] and Agustin Schiffrin[1,6]

[1]*Max-Planck-Institut für Quantenoptik, Hans-Kopfermann-Str. 1, D-85748 Garching, Germany*
[2]*Physik-Department, Technische Universität München, James-Franck-Str., D-85748 Garching, Germany*
[3]*Molecular Foundry, Lawrence Berkeley National Lab, 1 Cyclotron Road, Berkeley, California 94720*
[4]*Fritz-Haber-Institut der Max-Planck-Gesellschaft, Faradayweg 4-6, D-14195 Berlin, Germany*
[5]*Ludwig-Maximilians-Universität, Am Coulombwall 1, 85748 Garching, Germany*
[6]*School of Physics & Astronomy, Monash University, Clayton, Victoria 3800, Australia*

Nonlinear interactions between ultrashort optical waveforms and solids can be used to induce and steer electric current on a femtosecond (fs) timescales, holding promise for electronic signal processing at PHz ($10^{15}$ Hz) frequencies [Nature 493, 70 (2013)]. So far, this approach has been limited to insulators, requiring extremely strong peak electric fields (> 1 V/Å) and intensities (> $10^{13}$ W/cm$^2$). Here, we show all-optical generation and control of directly measurable electric current in a semiconductor relevant for high-speed and high-power (opto)electronics, gallium nitride (GaN), within an optical cycle and on a timescale shorter than 2 fs, at intensities at least an order of magnitude lower than those required for dielectrics. Our approach opens the door to PHz electronics and metrology, applicable to low-power (non-amplified) laser pulses, and may lead to future applications in semiconductor and photonic integrated circuit technologies.

Modern electronics relies on the control of electric current in solids [1]. The faster currents can be switched on and off in a device, the higher its processing performance. State-of-the-art high electron mobility transistors [2] can achieve switching rates of ~1 THz. Rates of ~100 THz, can be attained in semiconductors exposed to ultrashort optical fields, via photoconductive switching [3] and ω–2ω coherent control [4-9]. Recent experiments have shown that near-PHz current generation and control can be achieved in dielectrics via interactions with the electric field of intense few-cycle laser pulses [10-12]. This effect – result of highly nonlinear phenomena within the asymptotic limit of interband tunneling [13, 14] – comes at the expense of very high applied fields (> 1 V/Å), which can be a limitation for applications.

Here, we demonstrate ultrafast, direct-field control of current at substantially lower fields, in a material with a smaller bandgap ($E_g$). We have chosen GaN ($E_g \approx 3.4$ eV), which has attracted much interest for applications in optoelectronics and high-frequency and high-power electronics due to its high electron mobility, mechanical stability and heat capacity [15]. We show that charge displacement results from the interference between multiphoton excitation channels [16] in presence of field-induced intraband carrier motion and dynamic screening of the optical field. With increasing intensity, we observe a gradual transition from the multiphoton to the tunneling regime, supporting a unified quantum-mechanical picture valid in both limits.

We exposed the (0001) surface of wurtzite GaN to the waveform-controlled, linearly polarized visible/near-infrared (VIS/NIR) few-cycle laser pulses previously used in the prototypical study on silica [10] (see Supplement 1, Section 1). The instantaneous optical electric field, $F_L(t)$, was measured by attosecond streaking [17] in a parallel experiment [Figs. 1(a) and 4(b)]. The field was applied parallel to the surface, i.e., perpendicular to the permanent polarization of wurtzite GaN along its *c*-axis [18]. The stabilized carrier-envelope phase (CEP), $\varphi_{CE}$, was adjusted by varying the propagation length $\Delta l$ inside a pair of fused silica wedges. We considered applied electric field peak amplitudes, $F_0$, up to 0.9 V/Å (cycle-averaged peak intensity, $I_0 \leq 10^{13}$ W/cm$^2$). Gold (Au) electrodes were patterned onto GaN, allowing for direct measurement of optically-induced charge displacements (i.e., time-integrated electric currents) in the material [Fig. 1(a) and Supplement 1, Section 2].

Figure 1(b) shows the CEP-dependent fraction $Q_P$ of the charge per pulse collected by the unbiased Au electrodes as a function of $\Delta l$ and $\Delta\varphi_{CE}$. Here, $F_L(t)$ was applied perpendicular to the electrodes, along the *x*-axis; Fig. 1(a). The measured signal $Q_P$ reverses its sign periodically with CEP. Inverting the optical field ($\Delta\varphi_{CE} = \pi$) reverses the direction of the measured charge displacement: the instantaneous electric field of the laser pulse generates and controls $Q_P$, similar to the case of an insulator [10].

We measured $Q_P(\Delta l)$ for Au-GaN-Au junctions with inter-electrode distances of 100 nm, 5 and 10 μm, at various field strengths. Within this electrode separation range, the maximum value of $Q_P$ was given for 5 μm [4 ± 0.1 A fs = (4 ± 0.1) × $10^{-15}$ Coulomb at $F_0 \sim 0.4$ V/Å; Fig. 1(b)]; for 100 nm and 10 μm it was 1 ± 0.1 and 2.8 ± 0.1 A fs, respectively (both at $F_0 \sim 0.8$ V/Å). This hints at an optimal inter-electrode distance. A quantitative analysis of the maximum of $Q_P$ as a function of size of irradiated semiconductor area is beyond the scope of this study.

Figure 1(c) shows the (normalized) CEP-optimized transferred charge, $Q_P^{(max)}$, as a function of $F_0$ and $I_0$, for junctions with 100 nm and 10 μm inter-electrode spacing. For $F_0 \leq 0.45$ V/Å, the experimental data are well approximated by $Q_P^{(max)} \propto F_0^5$, independently of the junction size; this scaling law breaks down at larger fields. In comparison, data for SiO$_2$ from Ref. [10] shows a significantly higher order of nonlinearity, and a breakdown of the power law scaling at a much stronger field ($F_0 \approx 1.7$ V/Å). Notably, for the same $F_0$ (e.g. ~ 0.9 V/Å), signals for GaN are at least two orders of magnitude larger than those for SiO$_2$.

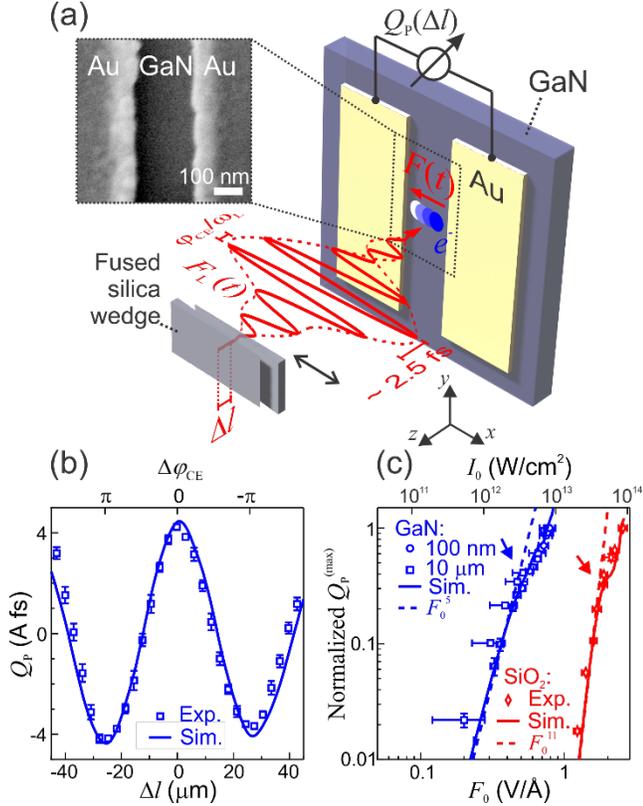

**Fig. 1.** (a) A (0001) wurtzite GaN surface patterned with gold electrodes (inset: SEM image) exposed to a CEP-controlled few-cycle VIS/NIR pulse with instantaneous electric field, $F_L(t)$. Electrodes are unbiased. (b) CEP-dependent component $Q_P$ of the collected charge per pulse as a function of propagation length $\Delta l$ in the fused silica wedges and of CEP change, $\Delta\varphi_{CE}$. Applied peak field amplitude, $F_0$: 0.4 V/Å. Inter-electrode spacing: 5 µm. (c) Maximum $Q_P$ [amplitude of sine fit of $Q_P(\Delta\varphi_{CE})$] as a function of $F_0$ and $I_0$ for 100 nm and 10 µm junctions. Data normalized with respect to values for maximum $F_0$. Data for SiO$_2$ [10] are shown for comparison. Arrows indicate breaking of the scaling power law. Solid curves: quantum-mechanical simulation.

When $F_0$ is varied, the transferred charge *shifts* with respect to $\Delta\varphi_{CE}$ [Figs. 2(b), (c)], i.e., the charge-balancing CEP, $\varphi_{CE}^{(+0)}$, for which $Q_P(\varphi_{CE}^{(+0)}) = 0$, increases monotonically [Fig. 2(a)]. Here, we focus (arbitrarily) on the charge-balancing CEP related to the rising edge of $Q_P(\Delta\varphi_{CE})$, i.e., $\partial Q_P(\varphi_{CE}^{(+0)})/\partial \varphi_{CE} > 0$. The dependence of $\varphi_{CE}^{(+0)}$ on $F_0$ is not affected by the inter-electrode separation; it is an intrinsic characteristic of the material, as evidenced by the comparison with the SiO$_2$ case [12]; Fig. 2a. The dependence $\varphi_{CE}^{(+0)}(F_0)$ allows for testing our theoretical model and aids the physical interpretation of our experiment.

Following the approach previously developed for SiO$_2$ [10], we decoupled *injection* (i) and *driving* (d) by exposing the junction to two synchronized, collinear VIS/NIR laser pulses with orthogonal electric fields $F_L^{(i)}(t)$ (parallel to electrodes, along y-axis; $F_0^{(i)} \approx 0.4$ V/Å) and $F_L^{(d)}(t)$ (perpendicular, along x-axis, $F_0^{(d)} \approx 0.06$ V/Å); Fig. 3(a). The CEPs $\varphi_{CE}^{(i)}$ and $\varphi_{CE}^{(d)}$ of the respective fields were set according to the inset in Fig. 3(b), i.e., such that $Q_P(\Delta\varphi_{CE}) = 0$ in single-pulse experiments (as in Figs. 1 and 2). Figure 4(b) shows $Q_P$ as a function of delay $\Delta t$ between the two pulses. For $\Delta t \approx 0$ fs, $Q_P(\Delta t)$ oscillates with a period of ~ 2.5 fs, i.e., the period of the optical field [Fig. 1(a)]. In Fig. 3(c), $\varphi_{CE}^{(d)}$ was changed by $\pi$; $F_L^{(d)}$ was reversed. Here, $Q_P(\Delta t)$ oscillates with the same period but is reversed in comparison to Fig. 3(b).

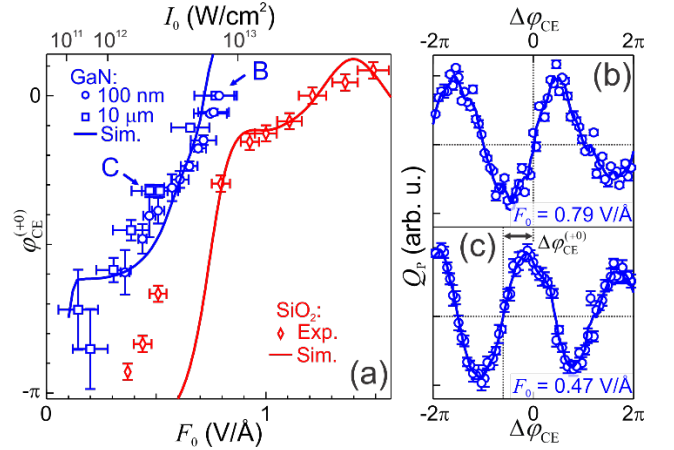

**Fig. 2.** (a) Charge-balancing CEP, $\varphi_{CE}^{(+0)}$, as a function of $F_0$ and $I_0$. Zero reference for $\varphi_{CE}^{(+0)}$ is set at $F_0 = 0.8$ V/Å. Data for SiO$_2$ are shown for comparison [12]. Solid curves: quantum mechanical simulation. (b), (c) $Q_P(\Delta\varphi_{CE})$ for $F_0 = 0.79$ [B in (a)] and 0.47 V/Å (C). Vertical dashed lines indicate the shift of $\varphi_{CE}^{(+0)}$ with $F_0$. Solid curves: smoothed experimental data.

To clarify the physical mechanism behind the generated current, we compared our experimental data with the results of quantum-mechanical (QM) simulations based on the numerical solution of the time-dependent Schrödinger equation [19] (Supplement 1, Sections 3–5). We considered optical transitions between three valence (VB) and two conduction bands (CB) for crystal momenta $k_x$ along the Γ–M direction of the Brillouin zone (BZ) [Figs. 4(a) and S1 in Supplement 1]. The electrodes orientation relative to the crystalline axes of the (0001) surface does not play an important role, since bands are isotropic in the vicinity of the Γ-point [20] and the considered field amplitudes are too low for most charge carriers to reach the BZ boundaries.

The theoretical curves (Figs. 1–3) are in good agreement with experiments within the full range of considered field strengths. We interpret our results as follows. The interaction between GaN and the laser pulse induces a nonequilibrium asymmetric population distribution in the VBs and CBs (Fig. 4a), leading to a CEP-dependent current along the optical field direction [5, 21]. This asymmetric population is due to quantum interference of excitation pathways [16, 21], which can be constructive for $k_x$ and destructive for $-k_x$, or *vice ver-* sa. This is shown in the calculated population distribution in Fig. S2c (Supplement 1), which is shifted from the BZ center and exhibits interference fringes. The interference of excitation pathways between electronic states in initial and final bands with energies $E_i(k)$ and $E_f(k)$ is determined by the accumulation of dynamic phase [22]

$$\Delta\phi_{fi}(k_x,t_1,t_2) = \frac{1}{\hbar}\int_{t_1}^{t_2} \Delta E_{fi}[K_x(t)]\,dt \quad (1)$$

due to intraband motion of electron-hole pairs between times $t_1$ and $t_2$ during exposure to the optical field. Here, $\Delta E_{fi}(k) = E_f(k) - E_i(k)$, $K_x(t) = k_x + eA(t)/\hbar$ is the semi-classical equation of intraband motion, $e > 0$ is the electron charge, and $A(t) = -\int_{-\infty}^{t} F(t')\,dt'$ is the vector potential of the total electric field $F(t)$ inside the Au-GaN-Au junction.

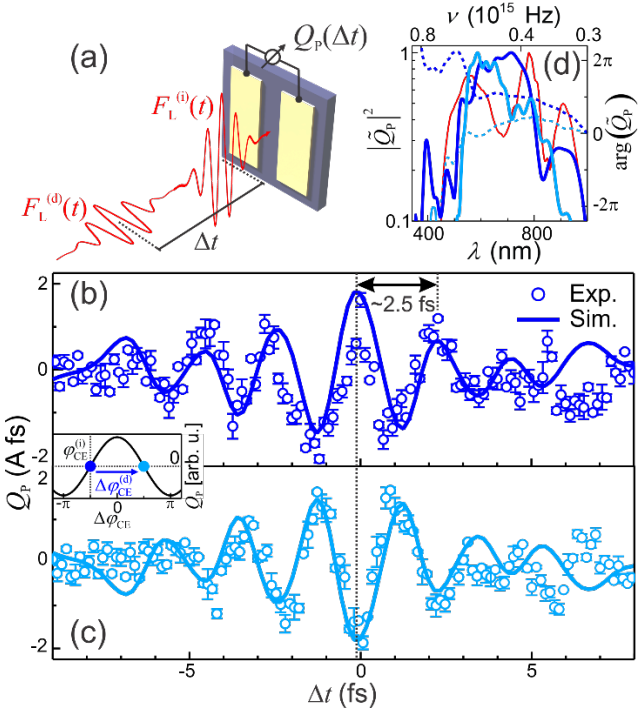

**Fig. 3.** (a) Injection-drive experiment. Two orthogonally-polarized VIS/NIR laser pulses, delayed by $\Delta t$, irradiate a 5 μm Au-GaN-Au junction ($F_0^{(i)} \approx 0.4$ V/Å; $F_0^{(d)} \approx 0.06$ V/Å). $\varphi_{CE}^{(i)}$ and $\varphi_{CE}^{(d)}$ are set such that $Q_P(\Delta\varphi_{CE}) = 0$ when $F_L^{(i)}(t)$ and $F_L^{(d)}(t)$ are applied independently. (b) CEP-dependent component $Q_P$ as a function of $\Delta t$. (c) Same as (b), with $\Delta\varphi_{CE}^{(d)} = \pi$. (d) Normalized modulus squared (solid) and phase (dashed) of the Fourier transform of the measured charge $\tilde{Q}_P = \mathcal{F}[Q_P(\Delta t)]$ in (b) (blue) and (c) (cyan), approximating the VIS/NIR pulse spectrum (red).

For example, if the dynamic phase accumulated over a laser cycle is a multiple of $2\pi$, $\Delta\phi_{fi}(k_x,t_1,t_1+2\pi/\omega_L) = 2\pi N$ ($N = 1, 2, …$), then contributions from different optical cycles interfere constructively, resulting in efficient excitation of electron-hole pairs. In the limit of a weak field, intraband motion can be neglected, $K_x(t) \approx k_x$, and $\Delta\phi_{fi} = 2\pi N$ yields the condition for absorbing $N$ photons: $\Delta E_{fi}(k_x) = N\hbar\omega_L$. Resonances of different orders $N$ can exist within an intense broadband pulse, and their interference determines the asymmetry of reciprocal-space population distributions.

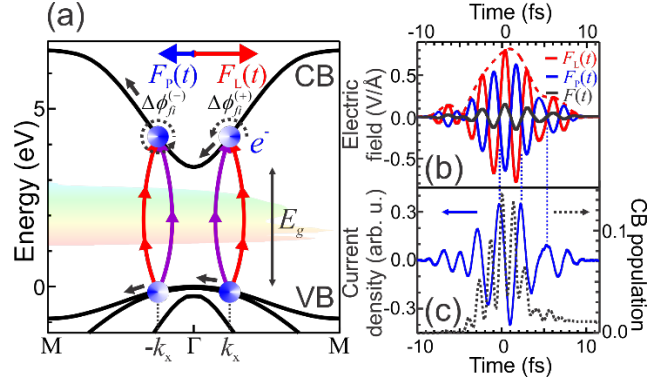

**Fig. 4.** (a) Mechanism of current injection in GaN. Charge carriers are created via interfering two- and three-photon transitions between valence (VB) and conduction (CB) band electronic states (blue circles: occupied; white: unoccupied). Heavy hole, light hole and crystal-field split-off VBs are shown. VIS/NIR pulse spectrum is depicted in the background. Dynamic phase shifts (dashed black arrows) $\Delta\phi_{fi}^{(\pm)} \equiv \Delta\phi_{fi}(\pm k_x, t_1, t_2)$ resulting from field-induced intraband motion of carriers (solid black arrows) determine whether interferences are constructive or destructive. (b) Applied optical electric field $F_L(t)$ (red), induced polarization field $F_P(t)$ (blue) calculated with quantum mechanical dynamic screening model, and total field $F(t) = F_L(t) + F_P(t)$. (c), Time-dependent current density $J(t)$ and electron population in the two lowest CBs calculated with quantum mechanical model including dynamic screening.

In the multiphoton regime, this quantum interference scenario yields a power-law scaling of the transferred charge $Q_P^{(max)} \propto F_0^{2N+1}$, where $N = \tilde{E}_g/\hbar\omega_L$ is the order of multiphoton interband transition and $\tilde{E}_g$ is the bandgap at a $k$-point where the corresponding multiphoton transition is allowed [16] (Supplement 1, Section 6). Notably, the scaling power law $Q_P^{(max)} \propto F_0^5$ ($N = 2$) observed in the experiment and in the QM model for $F_0 \leq 0.45$ V/Å [Fig. 1(c)] shows that, within this field magnitude range, the charge displacement is triggered by a multiphoton process. This observation is consistent with our estimation of the Keldysh [14] parameter $\gamma_K \gtrsim 2$ (Supplement 1, Section 6).

For stronger fields, the slope of $Q_P^{(max)}(F_0)$ decreases and diverges from the perturbative $F_0^{2N+1}$ scaling law. This is due to a combination of: (i) substantial screening of the external field by the optical-field-driven excited carrier displacement [23], and (ii) closing of the two-photon excitation channel [24] (Supplement 1, Section 6). The latter is a consequence of the ponderomotive energy becoming comparable to the photon energy, resulting in VB-to-CB transitions becoming non-resonant with the multiphoton process. This is indicative of nonperturbative dynamics and a gradual transition from the multiphoton to the tunneling regime.

The field amplitude dependence of the charge-balancing CEP, $\varphi_{CE}^{(+0)}$ (Fig. 2), allows to further test our QM model. The severe field-dependent shift of $\varphi_{CE}^{(+0)}$ is accurately reproduced by the theoretical curve. It is a direct consequence of the field screening due to the motion of charge carriers (Fig. S4 in Supplement 1).

The measured $Q_P(\Delta t)$ (Fig. 3) resolves the time-dependent oscillations of the optical field. Thus, it can be used for implementation of a solid-state attosecond streak camera (see Supplement 1, Section 7). The spectrum of $Q_P(\Delta t)$ extends to a maximal frequency of $f_{max} \sim 0.5$ PHz, closely resembling the pulse spectrum [Fig. 3(d)]. According to the cross-correlation theorem, the carrier injection associated with each optical cycle cannot be confined to a time window significantly broader than $1/(2 f_{max}) \sim 1$ fs, with significant contrast in the carrier excitation probability for adjacent optical cycles. The latter is ensured by the quasi-single-cycle character of the pulses and the nonlinearity of the process. Since the duration of the pulse is smaller than 4 fs, the current injection occurs in a time window smaller than 2 fs. This is consistent with the recently observed nonlinear ultrafast carrier excitation in semiconductors [25, 26].

In conclusion, we have demonstrated the injection and control of directly measurable current in a semiconductor (GaN) on a timescale shorter than 2 fs. Our observations highlight the interplay between interfering multiphoton excitation channels and intraband carrier motion. As the latter becomes more significant, deviations from the perturbative scaling law become more severe, as shown by our experimental and numerical data. This indicates a continuous transition from the multiphoton to the tunneling regime and emphasizes the role of dynamic screening of the optical field inside the solid. Our experiments pave the way for the development of ultrafast optically controlled solid-state electronics at intensities at least an order of magnitude smaller than those needed for an insulator. These intensities could further be decreased by optimization of junction geometries, opening the door to metrology for low-power (non-amplified) ultrashort laser pulse sources. Importantly, this approach would leverage and further expand existing semiconductor and integrated circuit technologies.

**Funding**. DFG Cluster of Excellence: Munich-Centre for Advanced Photonics; Max Planck Society; ERC consolidator grant AEDMOS; Integrated Initiative LASERLAB-Europe; Alexander von Humboldt Foundation; Swiss National Science Foundation; Marie Curie Fellowship (Project NANOULOP, No. 302157).

**Acknowledgment**. We thank Prof. F. Krausz for helpful discussions, F. Scholz and K. Forghani for providing the GaN samples, and B.D. Harteneck for help in sample design and fabrication.

# Sub-cycle optical control of current in a semiconductor: from the multiphoton to the tunneling regime. Supplementary material

Tim Paasch-Colberg[1], Stanislav Yu. Kruchinin[1], Özge Sağlam[2], Stefan Kapser[1], Stefano Cabrini[3], Sascha Muehlbrandt[1], Joachim Reichert[2], Johannes V. Barth[2], Ralph Ernstorfer[4], Reinhard Kienberger[2], Vladislav S. Yakovlev[1,5], Nicholas Karpowicz[1] and Agustin Schiffrin[1,6]

[1]*Max-Planck-Institut für Quantenoptik, Hans-Kopfermann-Str. 1, D-85748 Garching, Germany*
[2]*Physik-Department, Technische Universität München, James-Franck-Str., D-85748 Garching, Germany*
[3]*Molecular Foundry, Lawrence Berkeley National Lab, 1 Cyclotron Road, Berkeley, California 94720*
[4]*Fritz-Haber-Institut der Max-Planck-Gesellschaft, Faradayweg 4-6, D-14195 Berlin, Germany*
[5]*Ludwig-Maximilians-Universität, Am Coulombwall 1, 85748 Garching, Germany*
[6]*School of Physics & Astronomy, Monash University, Clayton, Victoria 3800, Australia*

## 1. Source of CEP-controlled few-cycle laser pulses

The laser pulses used in this work were generated with a customized titanium-sapphire chirped pulse amplifier (CPA) [1]. The optical electric field was linearly polarized. The pulse train had a repetition rate of 3 kHz. The pulses had a duration (full width at half maximum of the time-dependent cycle-averaged intensity) of ~ 3.8 fs and a central wavelength $\lambda_L \approx 760$ nm, which corresponds to a spectral range 450–1100 nm and central photon energy $\hbar\omega_L \approx 1.63$ eV. Maximum pulse energy was ~ 400 μJ. The laser beam was focused onto the Au-GaN-Au junction, resulting in waist sizes of ~ 50 μm, that is, significantly larger than the inter-electrode gap. The electric field peak amplitude $F_0 \leq 0.9$ V/Å and cycle-averaged peak intensity $I_0 \leq 10^{13}$ W/cm$^2$ were tuned by varying the aperture size of an iris, which does not affect the CEP at focus. The pulse carrier-envelope phase (CEP), $\varphi_{CE}$, was stabilized such that two consecutive pulses in the CPA output pulse train had a $\varphi_{CE}$ difference of $\pi$, that is, the CEP was modulated in the pulse train at half the laser repetition rate. The CEP was tuned according to $\Delta\varphi_{CE}(\Delta l) \propto 2\pi\Delta l\, \partial_\lambda n(\lambda_L)$, where $n(\lambda_L)$ is the wavelength-dependent refractive index of fused silica [2].

## 2. Patterning of (0001) wurtzite GaN with Au electrodes; measuring circuit

The photoactive material of the devices used in the measurements consisted of 300 nm thick, *n*-doped wurtzite GaN ($N_d = 10^{17}$ cm$^{-3}$) grown along the *c*-axis on a sapphire substrate [3]. Nano- and microscaled electrodes were deposited onto the (0001) GaN surface by electron-beam lithography. There is no anisotropic second-order electric susceptibility in the (0001) plane of this semiconductor [4]. In a first step, a thin layer of PMMA photoresist was spin-coated onto the surface [5]. The desired electrode geometry was obtained by developing the photoresist after exposure to the electron beam [6]. A 50 nm Au film was then deposited by electron beam physical vapor deposition [7] and the remaining PMMA was removed [see Fig. 1(a)]. Inter-electrode spacings of 100 nm, 5 μm and 10 μm were considered in this work. The electrodes were finally wire-bonded and connected to a high-gain current-voltage amplifier.

The output signal of the amplifier was Fourier filtered by a lock-in amplifier locked at half the repetition rate of the laser pulse train, allowing for direct measurement of the CEP-dependent fraction $Q_P$ of the total optically-induced charge displacement in the Au-GaN-Au junction. Experiments were performed in ambient conditions. Electrodes were not biased. Within the considered peak field range ($F_0 \leq 0.9$ V/Å), we did not observe sample damage. Each data point of $Q_P$ represents an average of 1500 measurements at a given $\Delta l$ (Figs. 1, 2 of main text) or $\Delta t$ (Fig. 3), with the standard deviation shown as vertical error bars. Horizontal error bars in Figs. 1(c) and 2(a) account for the fluctuations of pulse energy and beam waist at focus.

## 3. Quantum-mechanical model

Our quantum-mechanical simulations were based on the numerical solution of the time-dependent Schrödinger equation (TDSE) in the Houston basis [8]

$$\frac{d\alpha_i(k_x,t)}{dt} = \frac{i}{\hbar} eF(t)\sum_j X_{ij}[K_x(t)]\, \alpha_j(k_x,t) \\ \times \exp\left\{\frac{i}{\hbar}\int_{-\infty}^{t}\Delta E_{ij}[K_x(t')]\,dt'\right\}, \quad (S1)$$

where $|\alpha_i(k_x,t)|^2 = n_i(k_x,t)$ is the carrier population in the *i*-th band at crystal momentum $k_x$ and at time *t*, and $X_{ij}(k_x) = i\langle i,k_x|\partial_{k_x}|j,k_x\rangle$ is the optical matrix element between Bloch amplitudes.

We considered two CBs and three VBs on a one-dimensional (1D) *k*-space grid of 100 points. Our simulations included dynamic screening of the applied optical field due to the optical-field-induced displacement of excited charge carriers, which creates a total current density $J(t)$ (Fig. 4(c) of main text) and a time-dependent total macroscopic polarization $P(t)$ (Fig. S2b). The total electric field inside the solid, $F(t)$, was calculated self-consistently [9, 10]:

$$F(t) = F_L(t) + F_P(t) = F_L(t) - P(t)/\varepsilon_0, \quad (S2)$$

$$P(t) = \int_{-\infty}^{t} J(t')\,dt', \quad J(t) = \sum_{l\in VB} J_l(t), \quad (S3)$$

$$J_l(t) \propto -2e\,\mathrm{Re}\sum_{i,j}\int_{BZ}\frac{dk_x}{2\pi}\alpha_i^{(l)*}(k_x,t)\alpha_j^{(l)}(k_x,t) \\ \times \exp\left\{\frac{i}{\hbar}\int_{-\infty}^{t}\Delta E_{ij}[K_x(t')]\,dt'\right\}v_{ij}[K_x(t)]. \quad (S4)$$

Here, $J_l(t)$ describes the contribution to the current from all the bands in a calculation where an electron initially occupies valence band $l$ [11], $v_{ij}(k_x) = -i\hbar \langle i, k_x | \partial_x | j, k_x \rangle / m_0$ are the matrix elements of the velocity operator, and the factor of 2 accounts for spin degeneracy.

We modeled the two-pulse experiment (Fig. 3 of main text) by assuming that the *injecting pulse* $F_L^{(i)}$ polarized along the gap between the electrodes ($y$-axis) triggers charge carrier excitation and intraband motion parallel to the electrodes, and that the weak *driving pulse* $F_L^{(d)}$ induces intraband acceleration of electrons in the direction perpendicular to the electrodes ($x$-axis), but does not contribute to the carrier generation. The corresponding current density can be written as

$$J^{(d)}(t) \propto -2e\,\text{Re} \sum_{i,j,l} \iint_{BZ} \frac{dk_x dk_y}{(2\pi)^2} \alpha_i^{(i,l)*}(k_y,t)\alpha_j^{(i,l)}(k_y,t) \\ \times \exp\left\{\frac{i}{\hbar}\int_{-\infty}^{t} \Delta E_{ij}[K_y^{(i)}(t')]\,dt'\right\} v_{ij}^{(d)}[K_x^{(d)}(t)], \quad (S5)$$

where the probability amplitudes $\alpha_i^{(i,l)}(k_y,t)$ were obtained from numerical solution of the TDSE for the injecting pulse $F_L^{(i)}$, $v_{ij}^{(d)}$ and $K_x^{(d)}(t)$ were calculated along the driving field $F_L^{(d)}$. Fields inside the solid were calculated self-consistently as described above.

Proportionality coefficients for the calculated currents were chosen to correctly reproduce a linear response of GaN in the weak field limit [12]. The agreement between experiment and theory provides strong evidence that screening of the applied optical field by the charge displacement plays a crucial role in the observed phenomena.

In both the single-pulse (Figs. 1, 2, 4; main text) and two-pulse simulations (Fig. 3 of main text), the transferred charge $Q_P$ was calculated as an average of the residual polarization $\bar{P}$ along the drive field direction, at a fixed time $\tau = 12$ fs after the maximum of laser pulse envelope

$$Q_P = \mathcal{A}\bar{P} = \frac{\mathcal{A}}{\Delta\tau}\int_{\tau-\Delta\tau/2}^{\tau+\Delta\tau/2} P(t')dt' \quad (S6)$$

where $\mathcal{A} \sim 9$ μm$^2$ is the effective cross section of the active volume and the average interval $\Delta\tau = 2$ fs is chosen to be much larger than the period of interband coherence oscillations (~ 0.57 fs, see Fig. S2b).

## 4. Band structure and Brillouin zone of wurtzite GaN

In our quantum-mechanical model we assumed that $F_L(t)$ is polarized along the x-axis, which corresponds to the Γ–M direction in the Brillouin zone (Fig. S1). We considered two lowest conduction bands (CBs): c1 (blue) and c2 (light green), and three highest valence bands (VBs): hh (red), lh (purple), and ch (light blue).

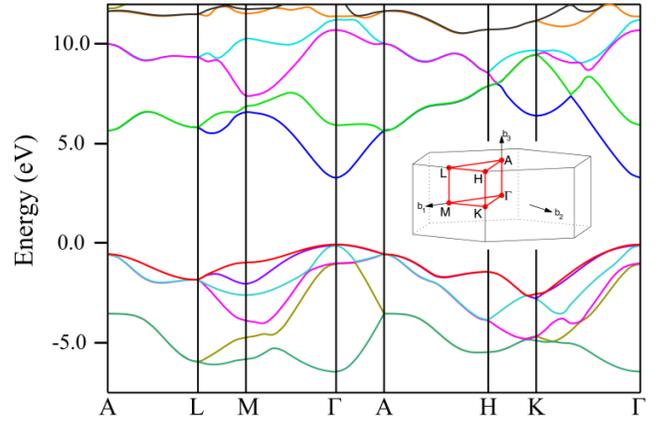

**Fig. S1.** Band structure of wurtzite GaN along the high-symmetry directions. Calculated with Wien2k package [13] using the TB09 meta-GGA exchange-correlation potential [14] with spin-orbit interaction. Inset: Brillouin zone of a wurtzite crystal and its irreducible part (adapted from Ref. [15]). The crystallographic *c*-axis is collinear to the $b_3$ vector (Γ–A direction), and the *a*-axis to $b_1$ (Γ–M direction).

## 5. Time-dependent polarization response and population distribution in GaN

Typical results for the single-pulse case obtained from numerical simulations with the dynamic screening model are shown on Fig. S2. Inset in Fig. S2b shows that after the pulse the total polarization oscillates around the value $\bar{P} \approx -0.005$ V/Å·$\varepsilon_0$, which determines the calculated charge, $Q_P \propto \bar{P}$.

The momentum-space CB population distribution calculated with our QM model (Fig. S2c) exhibits fringes due to quantum interference between multiphoton excitation pathways. The population asymmetry is the result of field-induced intraband motion influencing such interference. It should be noted that these momentum-space interference fringes (predicted here for charge carriers in a solid) are analogous to those observed for nonlinear photoionization of gas phase atoms [16].

## 6. Field dependence of transferred charge and charge-balancing phase shift

To further investigate the role of nonperturbative phenomena, we compared our QM simulations for the single-pulse experiment with results of a semiclassical (SC) two-band model, where the contribution of interband terms was neglected, and distributions of electrons and holes ($i = c_1$, lh) in the *k*-space were described by a Gaussian function

$$n_i(k_x,t) = n_0 \theta(t) \exp\left[-\frac{k_x^2}{2\sigma_{i,k_x}^2}\right], \quad (S7)$$

where $\theta(t)$ is the Heaviside unit step function, and the population amplitude $n_0 \propto F_0^4$ scales with the laser field amplitude according to perturbation theory [17], since two-photon absorption is the dominant excitation mechanism in this case ($E_g^{(GaN)} = 3.4$ eV, $\hbar\omega_L \approx 1.63$ eV).

The intraband motion of electron-hole wavepacket is considered using the Peierls substitution $k_x \to K_x(t)$ in the group velocity $v_i(k_x) = \partial_{k_x} E_i(k_x)/\hbar$. Thus, the current density in the SC model is defined as

$$J(t) \propto \sum_i \int_{BZ} e_i n_i(k_x,t) v_i[K_x(t)]\, dk_x, \quad (S8)$$

where $e_i = +e$ for VBs and $-e$ for CBs, $e > 0$ is the elementary charge.

We also provide a comparative study of two different field screening models: linear and dynamic. In the experiment, the laser beam spot size was much larger than the gap between electrodes, so that the field $F_L(t)$ was applied mostly to the electrodes. Therefore, in the first model, we assumed that screening is linear and determined by the well-known relation for the parallel-plate capacitor:

$$F(t) = \frac{F_L(t)}{\varepsilon}, \quad (S9)$$

where $\varepsilon \approx 5.57$ is the real part of the dielectric constant in wurtzite GaN at the central photon energy of the laser pulse [18], $\hbar\omega_L \approx 1.63$ eV.

This approximation fully ignores the nonlinear terms of interband polarization and contribution of free charge carriers. In the second model, the electric field inside the solid $F(t)$ was calculated dynamically [9, 12] from the time-dependent total macroscopic polarization $P(t)$, resulting from the optical-field-induced current density $J(t)$ in the material [see Methods in main text for details on the calculation of $J(t)$]

$$F(t) = F_L(t) + F_P(t) = F_L(t) - P(t)/\varepsilon_0, \quad (S10)$$

$$P(t) = \int_{-\infty}^{t} J(t')\, dt'. \quad (S11)$$

The normalized values of $Q_P^{(max)}$ obtained from all three models (QM with linear screening, QM with dynamic screening, and SC) are compared with the experimental data in Fig. S3. In the low-field limit (that is, at applied field amplitudes below $\sim 0.45$ V/Å), all three models converge and are in agreement with the experiment, following the $F_0^5$ law.

In the SC model, $F_0^4$ comes from the multiphoton excitation, $n_i(k_x,t) \propto F_0^4$, and an additional factor $F_0$ appears due to linearity of carriers group velocity with respect to the vector potential in the vicinity of $\Gamma$-point:

$$v_i[K_x(t)] \approx \frac{\hbar}{m_i} K_x(t) = \frac{\hbar}{m_i}\left[k_x + \frac{e}{\hbar} A(t)\right], \quad (S12)$$

where $m_i$ is the effective mass in band $i$.

The fact that $Q_P^{(max)} \propto F_0^5$ for $F_0 \leq 0.45$ V/Å represents compelling evidence that at low fields the observed physical phenomena can be interpreted perturbatively, within a scenario where VB electrons are excited to the CB via low-order, two- and three-photon absorption processes.

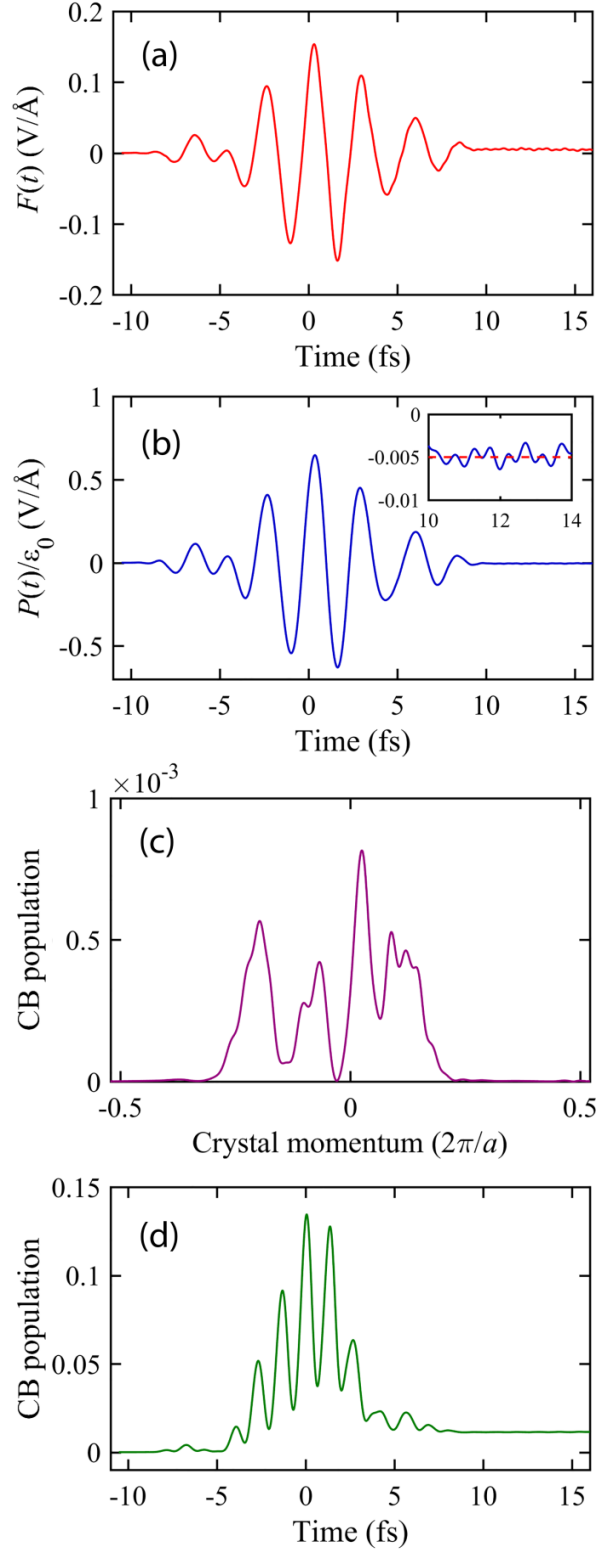

**Fig. S2.** Results of quantum-mechanical simulations with dynamic screening ($F_0 = 0.8$ V/Å, $\varphi_{CE} = 0$). (a) Total effective field $F(t)$ inside the Au-GaN-Au junction. (b) Total macroscopic polarization (solid line) and its average (dashed line). (c) Population distribution in the CBs after the pulse (at $t = 12$ fs) exhibiting interference fringes. (d) Time-dependent population of the CBs, integrated by $k_x$.

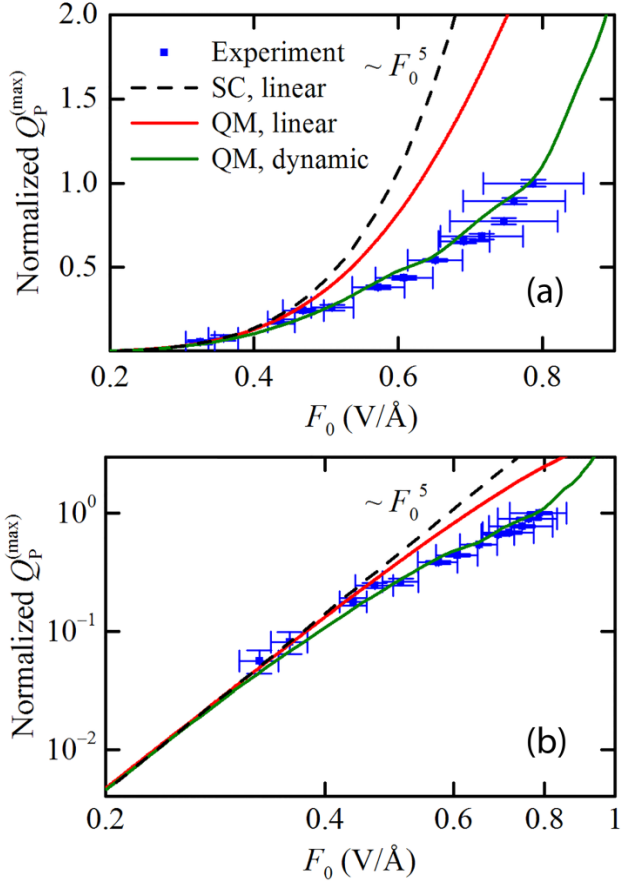

**Fig. S3.** CEP-optimized transferred charge $Q_P^{(\text{max})}$ as a function of laser field amplitude $F_0$ calculated with semiclassical (SC) and quantum-mechanical (QM) model. Comparison of simulations with experimental data for inter-electrode spacing 100 nm on a linear scale (a) and on a double logarithmic scale (b). Results of SC model with linear screening (dashed line) scale according to predictions of perturbation theory. The QM simulation with linear screening (red line) shows a slight decrease of the slope at higher fields, and the model with dynamic screening from the excited carriers (green line) agrees with experimental data in the full range of laser field amplitudes.

At applied field amplitudes above ~ 0.45 V/Å, experimental data and both quantum-mechanical models deviate from the $F_0^5$ dependence. Decrease of the slope in the model with linear screening (red curve in Fig. S3) can be explained by *multiphoton channel closing*. This effect was theoretically predicted by Keldysh for both atoms and solids [19], but has been experimentally investigated only in atomic systems [20, 21]. In strong fields, the ponderomotive energy $U_p$ increases and becomes comparable to the photon energy. Therefore, the effective energy of interband transition $\tilde{E}_g = E_g + U_p$ goes out of resonance with the lowest-order multiphoton process $N\hbar\omega_L \approx E_g$, which results in the decrease of its probability. The occurrence of multiphoton channel closing is indicative of a gradual transition from the multiphoton to the tunneling regime.

We clearly see from Fig. S3 that only the multiphoton channel closing effect is insufficient for explaining the decrease of the slope of the experimental data. Agreement between experiment and quantum-mechanical model — within the full range of considered field amplitudes — is only achieved when the screening field is calculated self-consistently with the TDSE (green curve in Fig. S3). Screening results from the displacement of free carriers by the applied optical field, reducing the effective electric field inside the solid. This leads to a further decrease of the transferred charge slope.

Additional information on the strong-field dynamics can be obtained from the dependence of charge-balancing phase $\varphi_{\text{CE}}^{(+0)}$ on the field amplitude. Fig. S4 shows that, at applied field strengths > 0.45 V/Å, the model with linear screening again diverges from experimental data with $\varphi_{\text{CE}}^{(+0)}(F_0)$ remaining practically constant. The simulation with dynamic screening reproduces the experiment, showing that the field-dependent phase shift is due to superposition of the applied laser field with the strongly nonlinear polarization field induced by excited and displaced charge carriers.

The $F_0$-dependent shift of $\varphi_{\text{CE}}^{(+0)}$ can be understood as the consequence of hysteresis-like behavior of polarization field with respect to the applied optical field (Fig. S5b). It is caused by the fact that, when the applied field is strong enough to excite free charge carriers, the collective electron motion in the material, which gives rise to the polarization field $F_P(t)$, is not perfectly synchronized to $F_L(t)$. The hysteresis results in temporal delays between $F_P(t)$ and $F_L(t)$, which, as $F_0$ increases, become more significant, resulting in a shift of the CEP of $F_L(t)$ for which $Q_P$ is maximum.

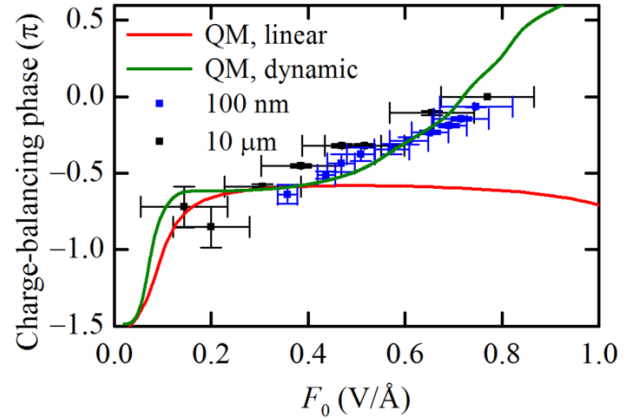

**Fig. S4.** Dependence of the charge-balancing phase $\varphi_{\text{CE}}^{(+0)}$ on laser field strength: linear versus dynamic screening. Simulation with linear screening shows phase stability at higher fields, since the effective field within the solid is proportional to the incident laser field. Dynamic screening model reproduces the experimentally observed field-dependent phase shift, providing evidence that the total effective field in the solid is significantly influenced by the laser-driven motion of free charge carriers.

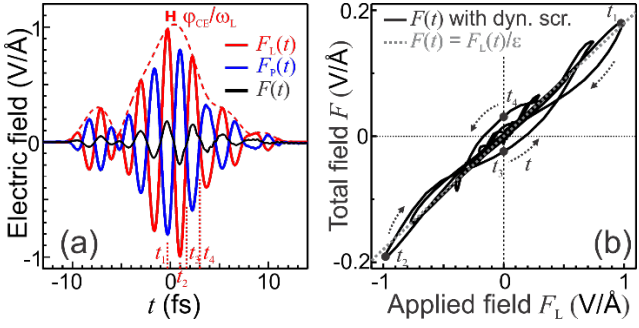

**Fig. S5.** Calculated dynamical polarization response of wurtzite GaN exposed to the optical field $F_L(t)$ ($F_0 = 1$ V/Å). (a) Polarization field $F_P(t) = -P(t)/\varepsilon_0$ and total effective field $F(t) = F_L(t) + F_P(t)$ in GaN. Times $t_1$, $t_2$, $t_3$ and $t_4$ indicate extrema and zero-crossings of $F_L(t)$. (b) Parametric plot of total effective field $F(t)$ vs. applied optical field $F_L(t)$. For optical cycles with maximum field amplitude (that is, for $t \geq t_1$), the polarization response in the material is delayed with respect to the applied field, causing hysteresis-like behavior. The linear response $F(t) = F_L(t)/\varepsilon$ is shown for comparison.

The multiphoton scenario discussed here leads to a physical picture that is intuitive in the perturbative regime, but less so for strong fields, where the number of interfering channels greatly increases. Moreover, the interpretation of the transferred charge in semiclassical terms, as a displacement of the electron-hole wavepacket, provides an intuitive explanation of the observed scaling laws, but neglects quantum interference of electron-hole wavepackets launched by different optical half-cycles and assumes that the laser-driven intraband motion of charge carriers is more important than interband excitation dynamics. It does not provide a self-consistent description of multiphoton transitions in the presence of significant intraband motion. Conversely, our QM model accounts for both quantum interference and intraband motion, which influence each other [see Eq. (1) in main text] and thus cannot be disentangled in general.

The QM model is in agreement with the two-pulse experiment in Fig. 3 of the main text. The strong injection field $F_L^{(i)}$, when applied independently, displaces carriers parallel to the electrodes, and no charge is collected by the circuit. The drive field $F_L^{(d)}$ is not strong enough to generate a measurable current by itself (Fig. 1(c) of main text). When both fields are applied, $\Delta t$ determines the instant of charge carrier injection with respect to $F_L^{(d)}$, and hence the net momentum change in the direction perpendicular to the electrodes. This is consistent with the observed reversing of $Q_P(\Delta t)$.

The oscillation of $Q_P(\Delta t)$ with the period of the laser field is similar to the case of $SiO_2$ [9, 12, 22]. However, the slightly reduced contrast of these oscillations for GaN could be explained by the smaller bandgap, and thus smaller degree of nonlinearity of the optical excitation (Fig. 1(c) in main text). In comparison with silica, the smaller GaN bandgap is also responsible for the substantially larger amount of excited charge carriers, which results in significant screening of the external field and hence severe shifts of $\varphi_{CE}^{(+0)}$ (Figs. S4 and 2 of main text).

It is also important to note that dynamical Bloch oscillations and Wannier-Stark localization [9, 23-28] are not expected to play a significant role in the charge carrier dynamics of GaN. Indeed, the maximum peak laser field $F_0$ to which GaN was exposed was below 1 V/Å, corresponding to a maximum effective peak field of ~ 0.2 V/Å inside the solid (taking screening into account). This is significantly smaller than estimated critical field $F_{crit} = E_g/(ea) \approx 1$ V/Å ($a \approx 3.2$ Å is the lattice constant of GaN in the (0001) plane), above which WSL becomes important. This is further justified by the values of the Keldysh parameter $\gamma_K \geq 2$ — indicative of multiphoton regime — and by the fact that most electrons do not reach the boundary of the first Brillouin zone during the interaction with the laser field. The Keldysh parameter is determined by $\gamma_K = \omega_L \sqrt{mE_g}/eF_{max}$, where $m = m_c m_{lh}/(m_c + m_{lh})$ is the reduced effective mass, $m_c \approx 0.22\, m_0$ and $m_{lh} \approx 0.18\, m_0$ are the effective masses in the first CB and in the light-hole band, respectively, and $F_{max} = F_0/\varepsilon(\omega_L)$ is the maximum amplitude of the total electric field in the Au-GaN-Au junction, with $\varepsilon(\omega_L) \approx 5.57$ for GaN. We only considered transitions between the light-hole band and the lowest CB since the corresponding reduced effective mass is the smallest.

Although our QM model and proposed physical scenario accurately reproduce and explain our experimental observations, they still leave room for further developments, especially in the observation and understanding of real-time sub-femtosecond dynamics of photoexcited charge carrier populations.

In particular, electronic dephasing and relaxation were not taken into account in our models. On the one hand, this is justified by the fact [29] that the measured dephasing time due to electron-phonon interactions in GaN is $T_2 \sim 30$ fs. This is much longer than the laser pulse duration, during which the optical-field-induced electron dynamics of interest takes place. On the other hand, dephasing of single-particle states due to electron-electron interaction can take place on timescales comparable to the VIS/NIR optical cycle. It is important to note, however, that dephasing does not modify the considered physical mechanisms of current injection (that is, interference between interband multiphoton transitions, and intraband carrier acceleration following two-photon excitation), but rather changes their relative contribution. These aspects regarding dephasing were established by performing additional numerical simulations based on the semiconductor Bloch equations (SBE) with different finite dephasing times. These simulations did not show a qualitative difference from the cases with infinite dephasing time, when the SBE are equivalent to the TDSE. Since our measurements do not allow directly accessing the real-time photoinduced carrier dynamics, a detailed investigation on the influence of dephasing and electron-electron interactions is beyond the scope of this work and will be subject to further studies.

## 7. Transferred charge in the two-pulse configuration

For the two-pulse experiment (Fig. 3 in the main text), we assumed that the weak driving field does not induce the interband transitions. Therefore, we can take into account only the intraband part of the current density

$$Q_\text{P}(\Delta t) = \iint_\mathcal{A} d\mathbf{a} \cdot \int_{-\infty}^{+\infty} dt\, \mathbf{J}^{(\text{d})}(t,\Delta t)$$
$$\approx 2\mathcal{A} \sum_i \int_{\text{BZ}} \frac{d^3 k}{(2\pi)^3} \int_{-\infty}^{+\infty} dt\, e_i n_i(\mathbf{k},t)\, \mathbf{e}_x \cdot \mathbf{v}_i[\mathbf{K}(t-\Delta t)], \quad \text{(S13)}$$

where $\mathbf{e}_x$ is the unit vector normal to the active surface area $\mathcal{A}$.

In the effective mass approximation, group velocity is given by (S12). Including the charge sign into the hole effective masses ($m_i \to -m_i$, $i \in \text{VB}$), one obtains

$$Q_\text{P}(\Delta t) \approx 2e\hbar\mathcal{A} \sum_i \int_{\text{BZ}} \frac{d^3 k}{(2\pi)^3}$$
$$\times \int_{-\infty}^{+\infty} dt\, \frac{n_i(\mathbf{k},t)}{m_i} \left[ k_x + \frac{e}{\hbar} A^{(\text{d})}(t-\Delta t) \right] \quad \text{(S14)}$$

The contribution of the first term in square brackets is negligible, because the population asymmetry due to interband transitions is induced by injecting field along perpendicular direction, so we finally get

$$Q_\text{P}(\Delta t) \propto \int_{-\infty}^{+\infty} dt\, n(t) A^{(\text{d})}(t-\Delta t), \quad \text{(S15)}$$

$$n(t) = 2e^2 \int_{\text{BZ}} \frac{d^3 k}{(2\pi)^3} \sum_i \frac{n_i(\mathbf{k},t)}{m_i}. \quad \text{(S16)}$$

Equation (S15) shows that within the effective mass approximation, $Q_\text{P}(\Delta t)$ can be expressed as the convolution between the vector potential of the weak drive field inside the solid, $A^{(\text{d})}(t) = -\int_{-\infty}^{t} F^{(\text{d})}(t')\, dt'$, and the carrier populations $n_i(\mathbf{k},t)$ induced by $F_\text{L}^{(\text{i})}$ at time $t$ in the $i$-th band. It reveals an analogy between our two-pulse measurements and attosecond streaking in vacuum [30, 31]. The main difference is that the free electron momentum $\mathbf{p}$ and mass $m_0$ are replaced by the crystal momentum $\hbar\mathbf{k}$ and the effective masses $m_i$ of the charge carriers in the populated bands. Deconvolution of this equation allows for resolving the real-time optical waveform inside the semiconductor, with sub-cycle resolution. Therefore, our technique implements a solid-state attosecond streak camera.